\documentclass[preprint]{iucr}              
\usepackage{graphicx}
\usepackage{rotating}
\usepackage{longtable}
\paperprodcode{a000000}      
\paperref{xx9999}            
\papertype{LN}               

\paperlang{english}          
\journalcode{J}              
\journalyr{2017}
\journaliss{1}
\journalvol{63}
\journalfirstpage{000}
\journallastpage{000}
\journalreceived{0 XXXXXXX 0000}
\journalaccepted{0 XXXXXXX 0000}
\journalonline{0 XXXXXXX 0000}

\begin{document}                  

\title{The modular SAXS data correction sequence for solids and dispersions.}
\shorttitle{SAXS data correction sequences}

\author[a]{B. R.}{Pauw}{brian.pauw@bam.de}
\author[b]{A. J.}{Smith}
\author[b]{T.}{Snow}
\author[b]{N. J.}{Terrill}
\author[a]{A. F.}{Th{\"{u}}nemann}

\aff[a]{Bundesanstalt f\"ur Materialforschung und -pr\"ufung (BAM), 12205 Berlin, Germany}

\aff[b]{Diamond Light Source Ltd., Diamond House, Harwell Science \& Innovation Campus, Didcot, Oxfordshire, OX11 0DE, \country{United Kingdom}}

\shortauthor{Pauw, Smith, Snow, Terrill, and Th{\"{u}}nemann}

\maketitle

\begin{synopsis}
A data correction sequence is presented, consisting of elementary steps, to extract the small-angle X-ray scattering cross-section from the original detector signal(s). It is applicable to solid samples as well as dispersions. 
\end{synopsis}

\begin{abstract}

Data correction is probably the least favourite activity amongst users experimenting with small-angle X-ray scattering (SAXS): if it is not done sufficiently well, this may become evident during the data analysis stage, necessitating the repetition of the data corrections from scratch. A recommended, comprehensive sequence of elementary data correction steps is presented here to alleviate the difficulties associated with data correction. When applied in the proposed order, the resulting data will provide a high degree of accuracy for both solid samples and dispersions. The solution here can be applied without modification to any pinhole-collimated instruments with photon-counting, direct detection area detectors. 

\textsc{keywords:} small-angle scattering, accuracy, methodology, data correction
\end{abstract}

\section{Introduction}

Attaining a high standard for data quality is paramount for any detailed analysis. This is of great importance for small-angle scattering in particular, where the largely featureless scattering patterns may easily be over- or under-fitted by an inexperienced user. Therefore, the provision of a consistent set of data corrections, which will also put well-founded uncertainty estimates on the resulting values, is a necessary addition to any small-angle scattering laboratory. Previous work on data correction procedures tended to follow an integral or ad-hoc approach, incorporating a limited subset of the available corrections, offering little flexibility or chances for tracing the effects of every individual correction \cite{Stothart-1987,Strunz-2000,Dreiss-2006}. A modular approach consisting of a sequence of elementary data correction steps would allow laboratories to select the subset of importance for their experiments or instruments, while allowing rapid evaluation of the significance. While most of the individual data correction steps that can be considered to achieve high quality data have been comprehensively collated before \cite{Pauw-2013a}, the recommended sequence in which they can be applied has not been published. This was due to a previous lack of software that might benefit from such a scheme, and because the sequence was still under development at the time. 

With the recent emergence of various comprehensive, modular data correction software packages \cite{Basham-2015,Filik-2017,Arnold-2014,silx,Benecke-2014,Nielsen-2009,Foxtrot,pySAXS}, establishing a recommended starting point for implementing such a data correction schema seems pertinent. This schema can be used as the core of a data correction software package, or as a reference correction sequence against which (faster) alternatives can be proven. It is hoped that adherence to this schema will improve the comparability of results obtained at different instruments. 

Over the last few years, the schema has been developed, tested and refined in practice, on both laboratory---as well as synchrotron---based SAXS instruments, with its modular nature making it easy to trace and verify the effect of each individual correction step on the detected signal and its uncertainties. In particular, this data correction scheme has been developed for modern instruments, and a direct-detection, photon-counting detector is highly recommended in order to achieve the best results. In this work, the use of a photon-counting detector is implicit, as the data correction steps necessary to compensate for the other detector types' inadequacies have been omitted for brevity.

Herein the schema will be presented, its individual correction's abbreviations briefly described, and the reasoning behind the placement will be clarified. 

\section{The Schema}

\begin{figure}
	\begin{center}
		\includegraphics[width=0.50\textwidth]{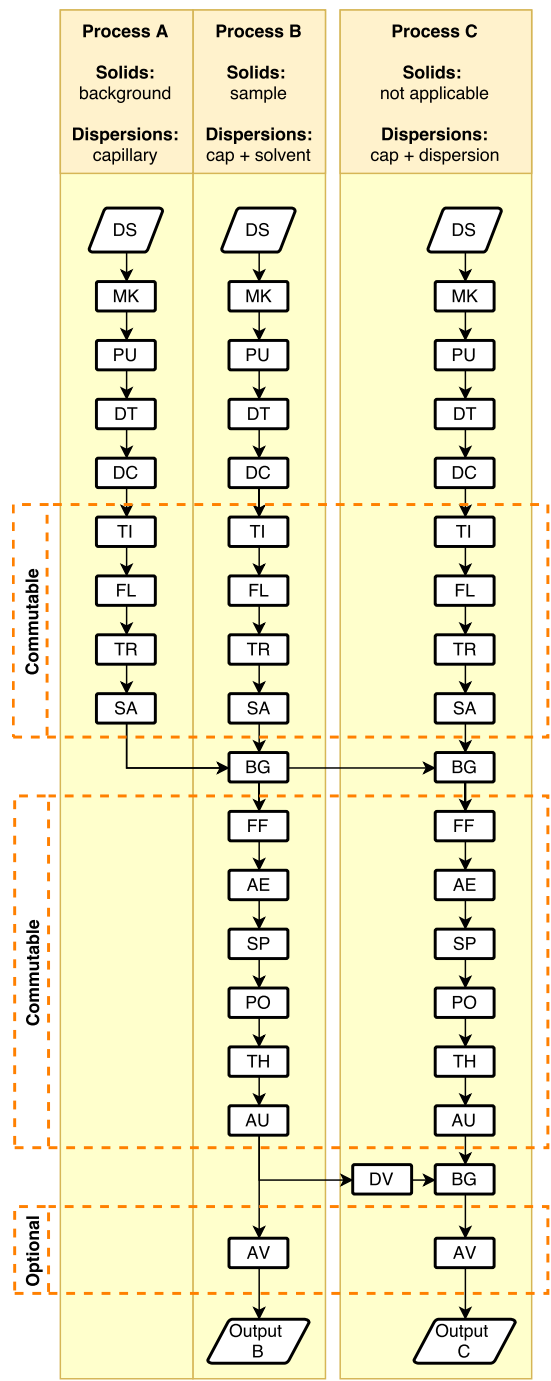}
		\caption{The recommended data correction sequence for solids (Process A \& B), or dispersions (Process A - C). Output B for solids is the corrected data in absolute units, for dispersions it is the solvent scattering in absolute units. Output C for dispersions is the sample scattering in absolute units. The azimuthal averaging step can be considered for isotropically scattering samples.}\label{fg:Schema}
	\end{center}
\end{figure}

The recommended data correction schema for solids and dispersions is presented in Figure \ref{fg:Schema}. For solid samples, this schema results in the scattering power of the solid in absolute units. For dispersions, both the solvent scattering as well as the sample scattering are obtained in absolute units. The solvent scattering can then be used for future samples by adding this to a solvent scattering library. The disadvantage of this approach is that the uncertainties of several operations are added twice to the output for dispersions (Output C), such as the uncertainties on the flatfield and polarization corrections. 


Note that for Process B or C for dispersions, the subtracted capillary signal used from Process A should be the same. That means that to obtain the solvent scattering cross-section, the same capillary should be used for Process A and B, and to obtain the dispersion scattering cross-section, the same capillary should be used for Process A and C. This, therefore, necessitates the use of reusable capillaries or flow-through cells.


\section{The Steps and Reasoning Behind the Sequence}

The mathematical expressions for each of the corrections below are described in \cite{Pauw-2013a}. Here, we focus on the justification of the steps and highlight the position dependency of some of them.

\begin{itemize}
\item{DS (Data Read-in):} Before starting any data corrections, the data must be read in correctly, where necessary compensating for the data storage peculiarities \cite{Knudsen-2013}. 
\item{MK (Masking):} Invalid pixels are masked so they are not considered in the following corrections.
\item{PU (Poisson Uncertainty Estimator):} The Poisson (counting) uncertainty needs to be calculated on the number of \emph{detected} photons, and therefore is carried out before the deadtime, darkcurrent or flatfield corrections.
\item{DT (Deadtime):} The signal is subsequently corrected for deadtime, returning the estimated number of photons arriving at each pixel based on the detected countrate.
\item{DC (Darkcurrent):} The subtraction of natural background radiation (including the steady flow of cosmic rays) forms the dominant component of the darkcurrent correction. With the aforementioned detector conditions, we should not see any significant contribution of the time-independent and flux-dependent darkcurrent components. 
\item{TI (Time):} A normalization to make the measurement independent of time. 
\item{FL (Flux):} A normalization to make the measurement independent of incident beam flux.
\item{TR (Transmission):} A scaling correction, correcting for the probability of absorption (and only absorption) within the sample. The transmission should, ideally be calculated by dividing the flux of the transmitted, scattered and diffracted signals by the incident flux. Note that the quality of the result is very strongly dependent on the quality of the transmission factor (in particular when the background subtraction operation is applied), and an accuracy of $>$99\% should be aimed for.
\item{SA (Self-absorption):} The sample self-absorption is the correction for the increased probability of scattered rays to be absorbed as they travel through slightly increased amounts of sample after the scattering event. This correction needs to be performed after the transmission correction: it represents a direction-dependent modification to the transmission correction, and does not replace the TR correction itself. It is feasible to implement and use for samples of plate-like geometry (only), with the plate surface perpendicular to the X-ray beam direction. It is related, and therefore placed next to the transmission correction.

\item{BG (Background subtraction):} The subtraction of the background signal is calculated only after the measurement-dependent corrections have taken place, as the various parameters (transmission, flux time, and therefore darkcurrent in particular) may differ.
\item{FF (Flatfield):} The flatfield correction, a multiplication matrix normalized to 1, corrects for interpixel sensitivity differences. This is the last of the corrections for detector imperfections.  
\item{AE (Angular Efficiency):} This correction compensates for variations in the detector efficiency depending on the photon angle of incidence onto the detector surface. It is detailed in Appendix \ref{ax:AE}. 
\item{SP (Solid-angle):} A (geometric) correction for the solid angle subtended by each pixel. This can be calculated on the basis of the instrument geometry alone. 
\item{PO (Polarization):} The polarization correction affects the probability of scattering events, both for polarized and unpolarized beams. In the latter case, it is a radially uniform (isotropic) correction.  The polarization correction is performed before the background subtraction for dispersions, so that older solvent measurements can still be used for correction. 
\item{TH (Thickness):} The thickness correction normalizes the data to units of reciprocal lengths. Note that the thickness used in this correction is the thickness of the solid sample or the liquid phase for dispersions only. A derivation for this is provided in Appendix \ref{ax:fBG}.
\item{AU (Absolute Units):} The absolute units correction scales the data to units of scattering cross-section, the fraction of radiation that is scattered per length of material per solid angle. This is commonly reported in units of $[\frac{\mathrm{d}\Sigma} {\mathrm{d}\Omega}] = m^{-1} sr^{-1}$ or $[\frac{\mathrm{d}\Sigma} {\mathrm{d}\Omega}] = cm^{-1} sr^{-1}$.

\item{DV (Displaced Volume):} This correction has not been included in the original work, but is described in Appendix \ref{ax:DV}. This correction can be done for dispersions with high volume fractions of analyte, but must be done on the solvent scattering signal only.  
\item{AV (Averaging):} This optional step reduces the dimensionality and size of the data, typically from 2D to a limited number of datapoints in 1D. This can be done azimuthally (to obtain $[\frac{\mathrm{d}\Sigma} {\mathrm{d}\Omega}]$ versus Q), or radially ($[\frac{\mathrm{d}\Sigma} {\mathrm{d}\Omega}]$ versus $\chi$). The azimuthal averaging is suitable for isotropic data, whereas the radial averaging is typically applied to anisotropic data, over a limited radial range, to extract degree of orientation; as is commonly utilised in fibre diffraction experiments.
\end{itemize}

\noindent The averaging from 2D to 1D are the last steps as 1) they are optional, and 2) the background subtraction process in particular can subtract anisotropic signals such as flares. In that case, the uncertainty is improved if the operation is done in 2D rather than after averaging.

Many of the corrections are multiplications, and therefore follow the law of commutation. The corrections that can be commutated have been grouped together where it is reasonable to do so such that a commutation would not affect the result. 
The commutability becomes clear when we write Process B as a pseudo-equation, with a $\rightarrow$ indicating a more involved operation, a $-$ indicating a subtraction, and a $\times$ indicating a multiplication operation with either a scalar or a vector:

\begin{eqnarray*}
		I(Q)\left[\frac{\mathrm{d}\Sigma} {\mathrm{d}\Omega}\right] &=& \left\{ \left[ (DS\rightarrow MK \rightarrow PU \rightarrow DT) - DC \right] \times TI \times FL \times TR \times SA - BG \right\} \\
        & & \times FF \times AE \times SP \times PO \times TH \times AU \rightarrow AV \\
\end{eqnarray*}

In order to further reduce the propagated uncertainties, the sequence for dispersions could be modified by postponing the flatfield, angular efficiency, polarization and solid-angle corrections in Process B and C, until after the second background subtraction in Process C is performed. This would reduce the uncertainties, as they are then only added once instead of twice (only for the sample, as opposed for both solvent and sample). Furthermore, if a flow-through cell is used, the thickness correction and absolute intensity scaling can also be postponed. While this seems desirable, the penalty is a drastic loss of generality: in this case the solvent scattering signal is no longer obtained in absolute units, thus reducing its value for future use. Conversely, in the recommended scheme, any stored solvent signal can be used for future data correction of an appropriate dispersion, significantly reducing overhead. It is, therefore, recommended to determine the flatfield, polarization and instrumental geometry with sufficient accuracy, so that the resultant increase in uncertainty can be kept to a minimum.

\section{A further practical modification}

In practice, the flux and transmission corrections can be combined. We define the transmission factor $T = \frac{I_1}{I_0}$, with the incident flux denoted as $I_0$, and the emergent flux (the sum of the transmitted, scattered and diffracted radiation) as $I_1$. Then, defining the prior detected signal $I_p(\bar{Q})$ and flux- and transmission-corrected signal $I_c(\bar{Q})$, we get:

\begin{equation}
I_c(\bar{Q}) = I_p(\bar{Q}) \frac{1}{I_0 T} = I_p(\bar{Q}) \frac{1} {I_1} 
\end{equation}

Combining these operations ostensibly negates the need for an upstream intensity monitor, to the great relief of many instrument scientists. However, as the transmission factor is still to be known for the self-absorption correction, their elation is likely to be short in duration.

\section{Instrumental effects for consideration in the analysis rather than the data corrections}

There are some effects which are, unfortunately, best considered in the scattering pattern analysis procedure rather than in the data correction procedure. There are three effects: The resolution function smearing, the multiple scattering effect, and the scattering length density contrast. We will discuss each of these briefly. 

The resolution effect originates from the uncertainty in the scattering vector for each individual photon. \emph{Some} of the origins of these uncertainties are well-defined, such as finite beam size and divergence, and the scattering vectors for an ensemble of photons will, therefore, exhibit a well-defined spread. This is known as the resolution function, and this can, in principle, be corrected for. The procedure to do this can be likened to a ``sharpening'' procedure in image processing, and carries the risk of introducing artifacts due to its ill-posed nature. As it is more prominent in the neutron scattering field, a workable solution has been developed already: the mathematically safer method for including the resolution contribution is to include the resolution function in the analysis. By convoluting, or ``smearing'' the model intensity with the resolution function, the problem is tractable, and can be taken into account without reservation \cite{Rennie-2013}.


The same holds for the multiple scattering contribution \cite{Warren-1966}. This is the probability that photons are scattered twice or multiple times, and is directly related to the scattering probability of a material for the energy used and its thickness. The multiple scattering contribution is hard to correct for in the original data. It is much easier to convolute the scattering pattern with the multiple scattering effect and likelihood, and to take it into account in that manner \cite{Rennie-2013}.

The last effect is the energy dependence of the scattering length density contrast. This energy dependence implies that, while the scattering vector is described independent of the energy, the scattering intensity will still be correlated, particularly near to absorption edges. There is, to our knowledge, no current solution for this, and information on the used energy must, therefore, always accompany a scattering curve.

\section{Conclusions}

We have presented a comprehensive data correction sequence, which can be used as the core of a software implementation, or as a reference correction sequence against which other, faster implementations can be proven. The sequence is chosen so that it returns useful information, in particular for dispersions, where both the absolute scattering signal from solvents as well as the analyte is obtained in as separate output signals. 

By presenting this schema, we hope to encourage unity and consistency in the worldwide data correction efforts, to the betterment of the small-angle X-ray scattering field.

\section{Acknowledgments}

The authors would like to thank the canSAS workshop attendees for their desire, support and encouragement that drove this work towards publication. We furthermore thank Steven Weigand for pointing us towards the DV correction, and Jacob Filik for showing us the AE correction. 


\referencelist[DataCorrectionSequence]

\appendix
\section{The Displaced Volume Correction: DV \label{ax:DV}}

The displaced volume correction is one that only needs to be considered for measurements on sample dispersions with a high volume fraction of analyte in the matrix. A rule of thumb would be to use this for analyte volume fractions of at least 1\%.

\begin{figure}
	\begin{center}
		\includegraphics[width=0.85\textwidth]{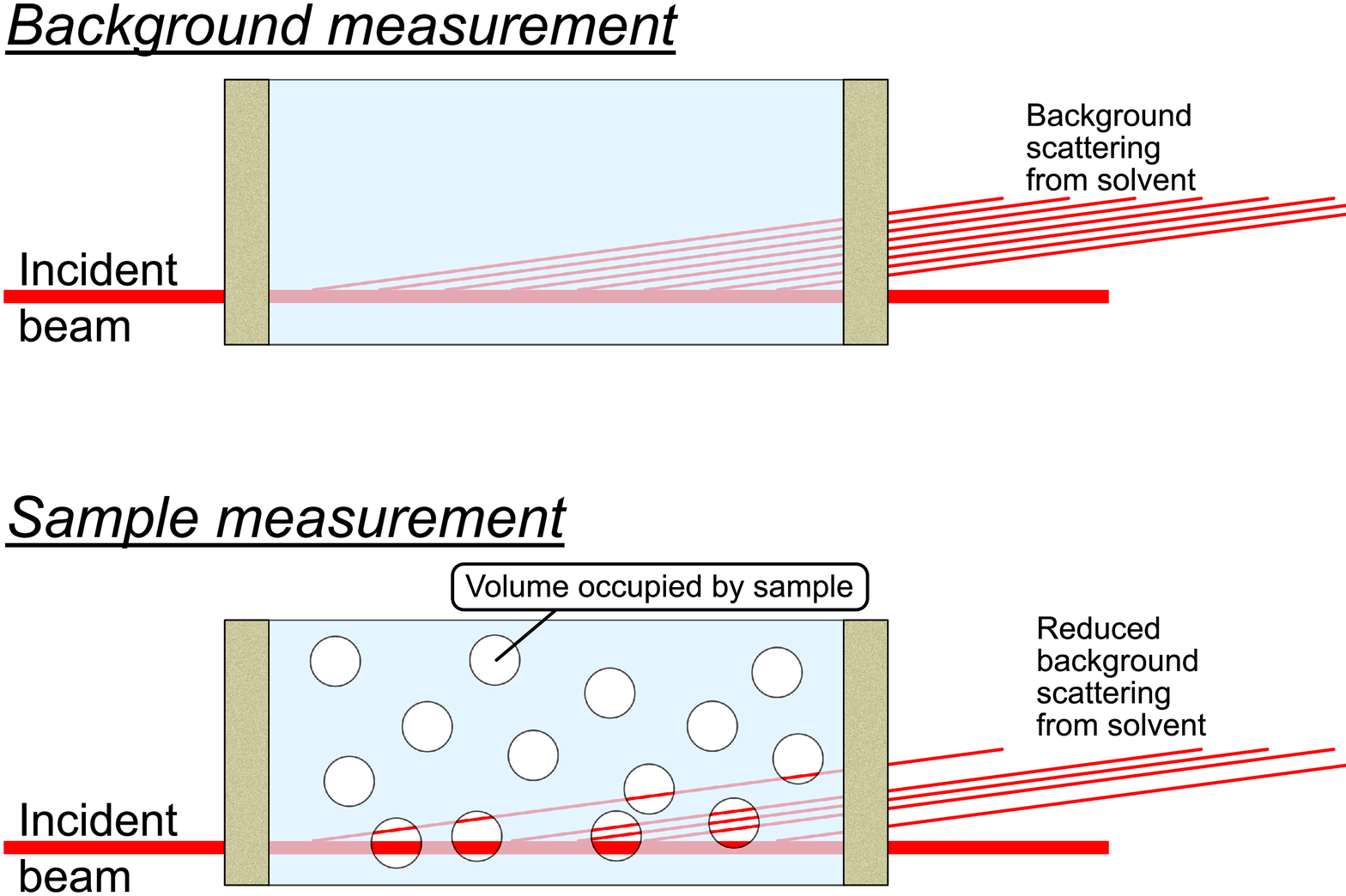}
		\caption{A schematic description of the displaced volume correction: a reduction in background signal when significant volume fractions of analyte are present. In that case, the thickness of background material is reduced, and its  contribution to the scattering signal reduces proportionally.}\label{fg:DVPrinciple}
	\end{center}
\end{figure}

In these cases there is a reduction in the amount of background material that the primary beam passes through, since a part of that space is now no longer occupied by the background material (see Figure \ref{fg:DVPrinciple}). In other words, there is a reduced length of background material in the beam, leading to a reduction in the background signal by an amount proportional to the volume fraction of sample in the beam. Perhaps counterintuitively, this is not something that is compensated for by the transmission measurement; the sample may have an identical overall absorption probability as the background, but still occupy a large fraction of the space.

What complicates matters is that this only reduces the background signal originating from the solvent, while leaving the background signal from the sample container walls unaffected. This means that the background signal needs to be “disassembled” into its components, and that the background scattering signal from the liquid needs to be reduced in a scaling procedure. For this reason, the schema in Figure \ref{fg:Schema} has two background subtractions, the first to separate both the solvent as well as the dispersion signal from the capillary walls, and the second to subtract the solvent signal from the solvent + analyte signal. Before applying the second subtraction, the solvent signal is multiplied with its (remaining) volume fraction. In this case, the scattering signal of the sample alone is obtained.

The second complication is that there is a bit of a chicken-and-egg problem; you cannot do this correction without knowledge on the volume fraction occupied by the sample. That volume fraction, however, may result from the scattering pattern analysis of the corrected scattering pattern (which you don’t have yet). It may be possible to do this correction in an iterative manner (an approach as yet untested). Alternatively, the volume fraction of analyte needs to be determined using other methods.

This correction will be significant, if: 1) the analyte volume fraction is significant, i.e. larger than 1 volume\%, and 2) the scattering signal from the sample is weak compared to the signal from the solvent. Proteins in solution and micellar systems are a prime example, but also dispersed polymers and vesicles may be affected.

\section{The Angle-dependent efficiency correction: AE \label{ax:AE}}

One additional correction can be considered, which takes into account the variation in detection probability of a photon passing through the detection layer at various angles \cite{Zaleski-1998}. When a photon passes through the detector at an angle perpendicular to the sensor surface, its detection probability is proportional to the absorption probability. This is, then, a function of the linear absorption coefficient (and thus the photon energy and sensor material), and the thickness of the sensor layer. If the photon were to pass through the detector layer at an angle, the amount of material it passes through is greater, and the detection probability increases\protect{\footnote{We ignore here, for simplicity, the fact that the detector surface is divided into individual pixels or voxels, each of which can absorb all or part of a photon's energy.}}. This means that the detection efficiency of a photon is greater when it arrives at oblique angles rather than perpendicular to the surface. 

\begin{figure}
	\begin{center}
		\includegraphics[width=0.75\textwidth]{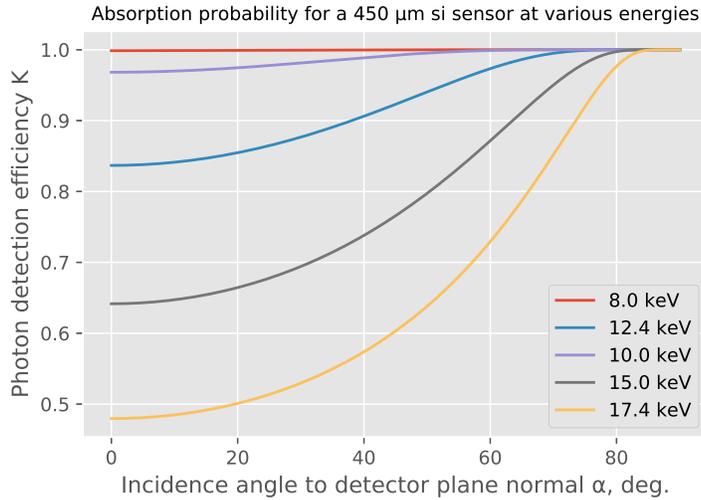}
		\caption{Detection efficiency of photons of various energies, dependent on their angle of incidence to the detector plane}\label{fg:KvsE}
	\end{center}
\end{figure}

This angle-dependent efficiency correction could be considered part of the flat-field response correction of the detector. Its source, however, is not due to detector imperfections, but lies in the instrument geometry coupled with the detector sensor thickness, and can, therefore, be considered separate. Since its magnitude can be easily estimated, it is straightforward to take it into account. If we rewrite the derivation from \citeasnoun{Zaleski-1998} to let $K$ represent the mass energy-absorption efficiency of a detector surface of thickness $d$ as a function of the angle of incidence $\alpha$ of a photon to the detector surface normal, we get:

\begin{equation}
K = 1 - \exp{ \frac{-\mu_{en} d} {\cos(\alpha)}}
\end{equation}

Where $\mu_{en}$ is the mass energy-absorption coefficient for silicon for a given energy. The correction of the observed intensity $I_{obs}(\textbf{Q})$ to the corrected intensity $I_{corr}$ then becomes: 

\begin{equation}
I_{corr}(\textbf{Q}) = I_{obs}(\textbf{Q}) / K
\end{equation}

The magnitude of this correction becomes larger with increasing energy, thinner detector surfaces, and increased angular coverage of the detectors. Figure \ref{fg:KvsE} shows the magnitude for various energies for a typical sensor thickness of 450 $\mu$m. Its magnitude may not be large for SAXS experiments, but it is easy to implement and correct for. Furthermore, when combining SAXS with WAXS data, the correction becomes more important.

\section{Deriving the Background Subtraction Sequence \label{ax:fBG}}

Dispersions are often measured inside a sample container (indeed, it is hard to do otherwise). This implies that we have an absorbing upstream and a downstream container wall, between which we have a particular length of sample, which also absorbs. Here, a derivation is shown to extract the sample scattering in this geometry, which forms a basis for the data correction sequence in the manuscript. For this calculation, the self-absorption correction of the scattered radiation is not considered. 

\subsection{Base definitions}

\emph{[note that the definitions made herein are for this appendix only, and do not apply to the general manuscript.]}

\subsubsection{System definitions}

The scattering system is considered to consist of a three-component, sandwich-like structure; an upstream sample container wall, followed by a sample, followed by the downstream sample container wall. All components are considered to be plate-like in shape, with the plate normal parallel to the direct beam. Furthermore, the distance between the sample and the detector is considered to be much larger than the thickness of the sample.

\subsubsection{Geometric definitions}

\begin{figure}
	\begin{center}
		\includegraphics[width=0.85\textwidth]{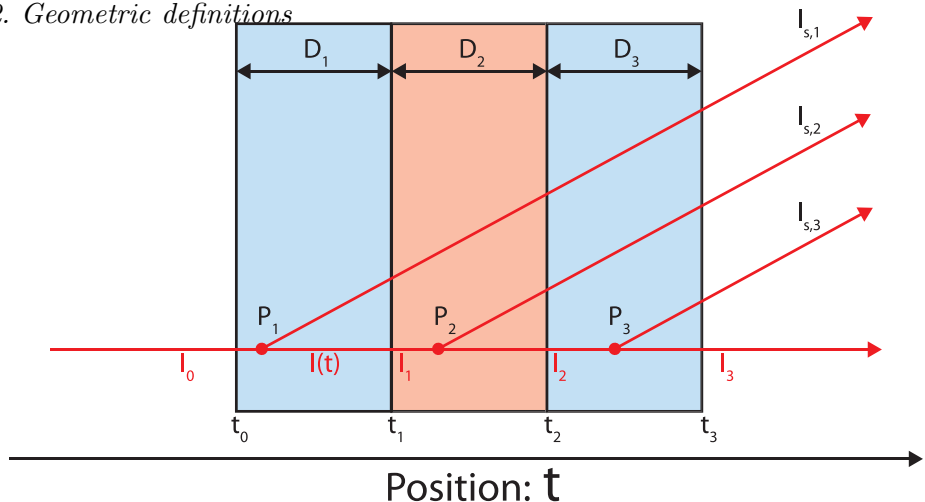}
		\caption{Schematic overview of the definitions used in the derivation of the background correction for dispersions ``sandwiched'' between two container walls.}\label{fg:fBGSchematic}
	\end{center}
\end{figure}

The upstream sample container wall is denoted by the subscript $_1$, the sample by $_2$, and the downstream sample container wall by $_3$. The following definitions are made (c.f. Figure \ref{fg:fBGSchematic}):

\begin{itemize}
\item{$D$}: The thickness of a phase
\item{$t$}: The running variable of distance traveled through all phases
\item{$t_0$}: position at the start of the upstream sample container component
\item{$t_1$}: position at the start of the sample component (end of the upstream sample container component)
\item{$t_2$}: position at the start of the downstream sample container component (end of sample component)
\item{$t_3$}: position at the end of the downstream sample container component
\item{$P_n(2\theta)$}: The scattering probability of phase $n$
\item{$I_0(t)$}: The primary beam intensity at position t
\item{$I_s(t)$}: The scattered intensity at position t
\item{$I_0$}: The primary beam intensity
\item{$I_1$}: The primary beam intensity entering the sample phase
\item{$I_2$}: The primary beam intensity entering the downstream sample container component
\item{$I_3$}: The primary beam intensity after absorption through all of the components
\item{$\mu_{n}$}: The linear absorption coefficient of phase $n$
\item{$2\theta$}: The angle of the scattered radiation
\item{$T_{n}$}: The transmission factor of a given phase or set of phases
\end{itemize}

\subsection{The derivation}
\subsubsection{Absorption of the unscattered beam}
X-ray absorption is defined as:
\begin{equation}\label{eq:Abs}
I_0(t) = I_0 \exp (-2\mu t)
\end{equation}
The beam intensities entering and exiting the various phases therefore work out as:
\begin{equation}\label{eq:basicAbs}
\begin{array}{rl}
I_1 =& I_0 \exp ( - \mu_1 D_1)\\
I_2 =& I_0 \exp \left[ - ( \mu_1 D_1 + \mu_2 D_2 ) \right]\\
I_3 =& I_0 \exp \left[ - ( \mu_1 D_1 + \mu_2 D_2 + \mu_3 D_3 ) \right]\\
\end{array}
\end{equation}

\subsubsection{Absorption of scattered beam by subsequent components}
Components in place after a scattered photon will absorb the scattered radiation with an absorption length slightly larger than the unscattered beam. The length of travel of the photon though subsequent materials is defined as:
\begin{equation}\label{eq:extraLength}
l = \frac{D - t}{\cos(2\theta)}
\end{equation}
The transmission factor $T$ of scattered radiation through subsequent phases therefore is:
\begin{equation}\label{eq:extraAbs}
T_n = \frac{I_{s,n}}{I_{s,n-1}} =  \exp ( - \mu_n l )
\end{equation}

\subsubsection{Intensity of the scattered beam in the scattering component}
The derivation of the scattered intensity, and direction-dependent transmission factor has been derived elsewhere \cite{Pauw-2013a}, where it was found to be:
\begin{equation}\label{eq:scatteringcomp}
I_1(t) = P_n I_0 \exp \left\{ -\frac{\mu}{\cos(2\theta)}\left[ t \cos(2\theta) + (D - t) \right] \right\}
\end{equation}

For the initial derivation, however, we do not consider the scattering angle-dependent increase in material pathlength, so that the term $\cos(2\theta) = 1$.

\subsubsection{Intensity scattered from component phases}
The scattered intensities of the individual components are defined as follows:
\begin{equation}\label{eq:componentI}
I_{sn} = I_{n-1}\int_{t_n}^{t^{n+1}} \exp (- \mu_n t) P_n \exp \left( - \mu_n \frac{D_n - t}{\cos(2\theta)} \right)  \mathrm{d}t  
\end{equation}

With $\cos(2\theta) = 1$, this simplifies to:
\begin{equation}\label{eq:componentIR}
\begin{array}{rl}
I_{sn} =& I_{n-1} P_n \exp \left( - \mu_n D_n \right) \int_{t_n}^{t^{n+1}} 1 \mathrm{d}t \\
= & I_{n-1} P_n D_n \exp \left( - \mu_n D_n \right) \\
\end{array}
\end{equation}

\subsubsection{Intensity scattered from the total}
The total scattered intensity is the sum of the scattering from all three components in the beam, attenuated by their subsequent phases. 
\begin{equation}\label{eq:totalI}
I_s = I_{s1} T_2 T_3 + I_{s2} T_3 + I_{s3}
\end{equation}

Substituting the components of equation \ref{eq:totalI} by with equations from \ref{eq:componentIR}, \ref{eq:basicAbs} and \ref{eq:extraAbs}, we get for the total scattered intensity of both sandwich-cell walls and the intermediate sample:
\begin{equation}\label{eq:full1}
\begin{array}{rl}
I_s = & I_0 P_1 D_1 \exp ( - \mu_1 D_1 )  \exp ( - \mu_2 D_2 ) \exp ( - \mu_3 D_3 ) \\
	& + I_1 P_2 D_2 \exp ( - \mu_2 D_2 ) \exp ( - \mu_3 D_3 ) \\ 
	& + I_2 P_3 D_3 \exp ( - \mu_3 D_3 ) \\
      = & I_0 P_1 D_1 \exp \left[ - (\mu_1 D_1 + \mu_2 D_2 + \mu_3 D_3) \right]  \\
	& + I_0 \exp ( - \mu_1 D_1 ) P_2 D_2 \exp \left[ - (\mu_2 D_2 + \mu_3 D_3 ) \right] \\ 
	& + I_0 \exp \left[ - (\mu_1 D_1 + \mu_2 D_2 ) \right] P_3 D_3 \exp ( - \mu_3 D_3 ) \\
      = & I_0 P_1 D_1 \exp \left[ - (\mu_1 D_1 + \mu_2 D_2 + \mu_3 D_3) \right]  \\
	& + I_0 P_2 D_2 \exp \left[ - (\mu_1 D_1 + \mu_2 D_2 + \mu_3 D_3 ) \right] \\ 
	& + I_0 P_3 D_3 \exp \left[ - (\mu_1 D_1 + \mu_2 D_2 + \mu_3 D_3 ) \right] \\
      = & I_0 \exp \left[ - (\mu_1 D_1 + \mu_2 D_2 + \mu_3 D_3) \right] (P_1 D_1 + P_2 D_2 + P_3 D_3)  \\
\end{array}
\end{equation}

Assuming phases 1 and 3 are identical, this simplifies to:

\begin{equation}\label{eq:full2}
\begin{array}{rl}
I_s = & I_0 \exp \left[ - (2 \mu_1 D_1 + \mu_2 D_2 ) \right] (2 P_1 D_1 + P_2 D_2 )  \\
\end{array}
\end{equation}

\subsubsection{Determining $P_1$}
Before we can continue, we must find out how to determine $P_1$. We do this in a background measurement, by measuring the scattering from the empty cell $I_b$ (in practice, the cell is ideally drawn to a vacuum, although the signal from air is assumed to be negligible). This implies that $P_2$ and $\mu_2$ are both zero as this phase is not present in the measurement. We then obtain $P_1$ from Equation \ref{eq:full1}:

\begin{equation}\label{eq:bgnd1}
I_b = 2 P_1 D_1 I_0 \exp ( - 2 \mu_1 D_1 ) 
\end{equation}
(NB: The first factor 2 originates from considering the upstream and downstream walls separately) 
So that: 
\begin{equation}\label{eq:bgnd2}
P_1 = \frac{I_b} {2 D_1 I_0 \exp ( - 2 \mu_1 D_1 ) } 
\end{equation}

\subsubsection{Extracting $P_2$}

Finally, we want to find the scattering probability of phase 2 $P_2$ (which is what we are really after), by rearranging equation \ref{eq:full2}:
 
\begin{equation}\label{eq:full3}
\begin{array}{rl}
2 P_1 D_1 + P_2 D_2  =& \frac{ I_s }{ I_0 \exp \left[ - (2 \mu_1 D_1 + \mu_2 D_2 ) \right] } \\
P_2 =& \frac{1}{D_2} \left\{ \frac{ I_s }{ I_0 \exp \left[ - (2 \mu_1 D_1 + \mu_2 D_2 ) \right] } - 2 P_1 D_1 \right\} \\
       =& \frac{1}{D_2} \left\{ \frac{ I_s }{ I_0 \exp \left[ - (2 \mu_1 D_1 + \mu_2 D_2 ) \right] }  - \frac{I_b} { I_0 \exp ( - 2 \mu_1 D_1 ) } \right\} \\
\end{array}
\end{equation}

Substituting the transmission factors for the empty cell $T_{\mathrm{1}} = \exp ( - 2 \mu_1 D_1 ) $ and cell plus sample $T_{\mathrm{1+2}} = \exp \left[ - (2 \mu_1 D_1 + \mu_2 D_2 ) \right] $, we can see that we arrive at the (more or less) standard background subtraction calculation:

\begin{equation}\label{eq:bgnd3}
P_2 = \frac{1}{D_2} \left\{ \frac{ I_s }{ I_0 T_{\mathrm{1+2}} }  - \frac{I_b} { I_0 T_{\mathrm{1}} } \right\} \\
\end{equation}
So, after this work, we find out that even when we thoroughly consider the scattering process of a sample sandwiched between two sample cell walls, we arrive at a simple equation for determining the sample scattering probability from the total measured intensity. 

\subsection{Final remarks}
There are interesting aspects when we use this background subtraction equation in practice. Firstly, we find that it is not necessary to determine the sample cell wall thickness $D_1$. Secondly, both the sample measurement and the background measurement are normalized to the thickness of the sample phase $D_2$ only. Lastly, it should be noted that this is, of course, only valid if the same sample cell is used for both the background and the sample measurement. 

Equation \ref{eq:bgnd3} as derived thus is represented using the modular data corrections as shown in Figure \ref{fg:Schema}. The thickness correction occurs after background subtraction, and the transmission and incident flux corrections have been applied before subtraction. The same background equation also works for simpler cases, for example when measuring a solid sample with an empty background.

\end{document}